\documentclass[10pt]{article}
\usepackage[latin9]{inputenc}
\usepackage{amsmath}
\usepackage{amssymb}
\usepackage{graphicx}
\usepackage{esint}
\usepackage[unicode=true,pdfusetitle,
 bookmarks=true,bookmarksnumbered=false,bookmarksopen=false,
 breaklinks=false,pdfborder={0 0 1},backref=section,colorlinks=false]
 {hyperref}
\usepackage{breakurl}

\makeatletter
\@ifundefined{date}{}{\date{}}
\usepackage{cite}

\usepackage{lineno}

\usepackage{microtype}\DisableLigatures[f]{encoding = *, family = * }

\usepackage[labelfont=bf,labelsep=period,justification=raggedright]{caption}

\renewcommand{\@biblabel}[1]{\quad#1.}

\makeatother

\begin{document}

\begin{flushleft}
\textbf{\Large Efficient transmission of subthreshold signals in complex
networks of spiking neurons} 
\\
 Joaquin J. Torres$^{1,*}$, Irene Elices$^{1,**}$, J. Marro$^{1}$\\
\textbf{{1} Department of Electromagnetism and Physics of the Matter,
University of Granada, Granada, Spain}\\
\textbf{$\ast$ E-mail: Corresponding jtorres@onsager.ugr.es }\\
\textbf{$**$ Presently at Grupo de Neurocomputación Biológica, Dpto. de Ingeniería
Informática, Escuela Politécnica Superior, Universidad Autónoma de
Madrid, Madrid, Spain}
\par\end{flushleft}

\textbf{
}

\section*{Abstract}

We investigate the efficient transmission and processing of weak,
subthreshold signals in a realistic neural medium in the presence of
different levels of the underlying noise. Assuming Hebbian weights for
maximal synaptic conductances -- that naturally balances the network with
excitatory and inhibitory synapses -- and considering short-term synaptic
plasticity affecting such conductances, we found different dynamic phases in
the system. This includes a memory phase where population of neurons remain
synchronized, an oscillatory phase where transitions between different
synchronized populations of neurons appears and an asynchronous or noisy
phase. When a weak stimulus input is applied to each neuron, increasing the
level of noise in the medium we found an efficient transmission of such
stimuli around the transition and critical points separating different
phases for well-defined different levels of stochasticity in the system. We
proved that this intriguing phenomenon is quite robust, as it occurs in
different situations including several types of synaptic plasticity,
different type and number of stored patterns and diverse network topologies,
namely, diluted networks and complex topologies such as scale-free and
small-world networks. We conclude that the robustness of the phenomenon in
different realistic scenarios, including spiking neurons, short-term
synaptic plasticity and complex networks topologies, make very likely that
it could also occur in actual neural systems as recent psycho-physical
experiments suggest.

\section*{Author Summary}

In this work we have investigated how weak, subthreshold stimuli can be
efficiently processed in an auto-associative spiking neural network with
dynamic synapses, in the presence of noise. We have described different
non-equilibrium phases including a memory phase, an oscillatory phase, with
random transitions among memory attractors, and a non-memory phase. Our main
finding is that system is able to efficiently process the relevant signals
at several well defined levels of the underlying stochasticity in the
system, which are related with phase transitions points. In addition, we
have proved the robustness of this intriguing phenomenology in different
situations which include several types of long term and short-term synaptic
plasticity and different network structural properties. This makes very
likely that this phenomenology can appear also in actual neural systems.

\section*{Introduction}

Many physical systems present ambient and intrinsic fluctuations that often
are ignored in theoretical studies to obtain simple mean-field analytical
approaches. Nevertheless, these fluctuations may play a fundamental role in
natural systems. For instance, they may optimize signals propagation by
turning the medium into an excitable one -- e.g., the case of ionic
channel stochasticity in neurons that can affect the first spike latency 
\cite{ref1,ref2} or enhance signal propagation through different neuronal
layers \cite{ref3} --, originate order at macroscopic and mesoscopic levels 
\cite{JMS,orderjapos} or induce coherence between the intrinsic
dynamics of a system and some weak stimuli it receives, a phenomenon known
as \textit{stochastic resonance} (SR) (see for instance \cite{SRG} for a
review). Precisely, this intriguing phenomenon has attracted the interest of
the computational neuroscience community for its possible implications in
the complex processing of information in the brain \cite{ref4,jap,ref5,indioplosone13,DrosteFCN13}, or as a way to control specific
brain states \cite{obermayer13}. In fact, most neural systems naturally
include the main factors involved in stochastic resonance, namely, different
sources of intrinsic and external noise and complex nonlinear dynamic
processes associated, for instance, to neuron excitability and information
transmission through the synapses. Thus, several experimental, theoretical
and numerical studies concerning the efficient transmission of weak signals
in noisy neural systems have been reported recently \cite%
{ref4,jap,ref5,kra,JM,mejias,noise,Gio, indioplosone13,DrosteFCN13}.

Particularly interesting are recent works that show that, differently to
what happen in traditional SR phenomena, noisy neural systems can optimally
process relevant information at more than one level of the ambient noise 
\cite{mejias,Gio}. However, it is not clear yet what are the main factors
responsible for this new phenomenology. It has been reported, for instance,
that this new phenomenon can appear in single perceptrons with dynamic
synapses \cite{mejias} as a consequence of the complex interplay between
dynamic synapses and adaptive threshold mechanisms affecting neuron
excitability. Also, similar phenomena has been reported in binary networks
with scale-free topologies \cite{kra}, so that these resonances may emerge
as a consequence of topological disorder. Finally, very recently, it has
been reported that optimal processing of weak relevant signals at different
levels of the underlying noise also occurs in \emph{auto-associative}
networks of binary neurons with dynamic synapses \cite{Gio}. Most of these
studies, however, consider very simple models at the neuron, network and
synapse levels. This makes difficult to extrapolate their results and
conclusions to actual neural systems.

Here, we present a full computational study of how weak relevant
subthreshold signals can be processed by neural systems in a more realistic
scenario, that is, a complex auto-associative network of spiking neurons
with dynamic synapses. We consider a network of $N$ \emph{integrate and fire}
neurons, and assume a long-term synaptic plasticity mechanism, due to
Hebbian learning, affecting the synapses connecting the neurons. In order to
increase the biological relevance of our study we also consider different
types of \textit{short-term\ synaptic\ plasticity,} such as the so called 
\emph{short-term depression} and \emph{short-term facilitation.} These
mechanisms introduce synaptic changes at short time scales, as it is expected
to occur due to the existence of a limited amount of neurotransmitters at
each synapse that can be released after some presynaptic stimulus, which
needs some time to recover after this stimulus.

In addition, to mimic any source of intrinsic or external ambient noise that
can affect neuron dynamics and its excitability, we add to each neuron
dynamics a source of uncorrelated Gaussian noise together with a weak
stimulus. In this way, by controlling the intensity of noise we can monitor
the levels of the noise at which the processing of the weak stimulus by the
system is more efficient. For simplicity, we assume in most of the cases
reported here a form of sinusoidal signal for the stimulus, which simulates
some relevant information to be processed by neurons, but other type of more
realistic signals can also be considered with the same results (see Results
section).

To see how efficient this weak stimulus in each neuron is processed by the
system, one can compare, for instance, the coherence in time between the
mean network activity and the stimulus by means some information transfer
measurement as a function of the noise intensity. In general for low noise,
the activity of the system cannot correlate with the stimulus due to its
small amplitude. If the intensity of noise is increased sufficiently, then
the noise is able to enhance the stimulus temporal features in such a way
that the network activity can start to correlated with it and a peak of
information transfer will appear. Nevertheless, if the noise intensity is
too high, the network activity will be dominated by the noise preventing the
input stimulus from being detected by the system.

On the other hand, activity dependent synaptic mechanisms, such as
short-term depression and short-term facilitation, may be highly relevant in
signal detection in noisy environments and can play a main role, for
instance, in SR \cite{mejias,Gio}. These synaptic mechanisms may modify the
postsynaptic neural response in a nontrivial way. Synapses can present
short-term depression when the amount of neurotransmitters that are
available for release whenever an action potential arrives is limited, and
consequently, the synapse may not have time to recover them if the frequency
of arriving spikes is too high. On the contrary, short-term facilitation is
determined by the excess of calcium ions in the presynaptic terminal which
can increase the postsynaptic response under repetitive stimulation. Both
synaptic processes could interact with noise and some neuron excitability
and adaptive mechanisms to induce a strong influence during the processing
of relevant signals or stimulus in the brain. In particular, it has been
recently reported in single perceptrons and in network of binary neurons
that the complex interplay among these synaptic mechanisms allow for
efficient detection of weak signals at different levels of the underlying
noise \cite{mejias,Gio} and maintain coherence for a wide range of the
intensity of such noise \cite{jap}.

In this work, we demonstrate that these intriguing emergent phenomena appear
also during the processing of weak subthreshold signals in more realistic
neural media and in many different conditions. Therefore, it is highly
likely that also they may appear in the actual neural systems, where
different types of signals and stimulus are continuously processing in the
presence of different sources of intrinsic and external noise. Moreover, the
fact that the processing of weak subthreshold signals occurs at well defined
different levels of noise -- normally one relatively low and the other
relatively high -- can have strong implications concerning how the signal
features are being processed. This can be clearly depicted in the case of
more realistic Poissonian signals (see Results section)

To demonstrate the robustness of our findings, we performed a complete
analysis of the emergent phenomena changing many variables in our system.
This confirms that the same interesting phenomena emerges in all these
situations, including, for instance the case in which the number of neurons
in the network and the number of stored patterns is increased. In addition,
the phenomenon of interest also remains for non-symmetric stored patterns
provided that there is a phase of transitions between a high activity state
(up state) and a low activity state (down state) in the network activity.
Including short-time synaptic facilitation at the synapses, competing with
synaptic depression, causes also intriguing features. This includes a
dependency with the level of facilitation, of the level of noise at which
the subthreshold signals are processed and detected and an enhancing of the
detection quality for large facilitation. Finally, we checked the robustness
of our finding for more realistic network topologies, such as diluted
networks, and complex scale-free and small-world topologies confirming that
phenomenon is robust also in these cases.

\section*{Materials and Methods}

The system under study consists of a spiking network of $N$ integrate and
fire neurons interconnected each other. The membrane potential of the $i-th$
neuron then follows the dynamics

\begin{equation}
\tau_{m}\frac{dV_{i}(t)}{dt}=-V_{i}(t)+R_{m}I_{i}(t)\hspace{1cm}%
0<V_{i}(t)<V_{th},  \label{IF}
\end{equation}
where $\tau_{m}$ is the cell membrane time constant, $R_{m}$ is the membrane
resistance and $V_{th}$ is a voltage threshold for neuron firing. Thus, when
the input current $I_{i}(t)$ is such that depolarizes the membrane potential
until it reaches $V_{th}=10mV$ an action potential is generated. Then, the
membrane potential is reset to its resting value -- that for simplicity we
assume here to be zero -- during a \emph{refractory period} of $\tau_{ref}=5\
ms$. We can assign binary values $s_{i}=1,0$ to the state of the neurons,
depending if they have their membrane voltage above or below the voltage
firing threshold $V_{th}.$ Furthermore, we assume that synapses between
neurons are \emph{dynamic} and described by the Tsodyks-Markram model
introduced in \cite{PNASTSM}.

Within this framework, we consider that the total input current $I_{i}(t)$
in the equation (\ref{IF}) has four components, that is, $I_{i}(t)=I_{0}+I_{i}^{ext}+I_{i}^{syn}+D\zeta(t)$ where $I_{0}$ is a
constant input current. The second term $I_{i}^{ext}$ represents an external
input weak signal which encodes relevant information and for simplicity we
assume to be sinusoidal, that is 
\begin{equation}
I_{i}^{ext}=d_{s}\sin(2\pi f_{s}t),
\end{equation}
with frequency $f_{s}$ and a small amplitude $d_{s}$. The fourth component
of $I_{i}(t)$ is a noisy term that tries to mimic different sources of
intrinsic or external current fluctuations, where $\zeta(t)$ is a Gaussian
white noise of zero mean and variance $\sigma=1$, and $D$ is the noise
intensity. Finally, the third component $I_{i}^{syn}$ is the sum of all
synaptic currents generated at neuron $i$ from the arrival of presynaptic
spikes on its neighbors. Following the model of dynamic synapses in \cite%
{PNASTSM}, we describe the state of a given synapse $j$ by variables $y_{j}(t),$ $z_{j}(t)$ and $x_{j}(t)$ representing, respectively, the
fraction of \emph{neurotransmitters} in active, inactive and recovering
states. Within this framework, active neurotransmitters $y_{j}(t)$ are the
responsible for the generation of the postsynaptic response after the
incoming presynaptic spikes, and become inactive after a typical time $\tau_{in}\sim2-3ms.$ On the other hand, inactive neurotransmitters can
recover during a typical time $\tau_{rec}$ which is order of a half second
for typical pyramidal neurons \cite{PNASTSM}, a fact that induces \emph{%
short-term synaptic depression}. Recovered neurotransmitters become
immediately active with some probability ${\cal U}$ -- the so called
release probability -- every time a presynaptic spike arrives to the
synapses. In actual synapses, ${\cal U}$ can increases in time with a
typical time constant $\tau_{fac}$ -- due to some cellular biophysical
processes associated to the influx of calcium ions after the arrival of
presynaptic spikes -- which induces the so called \emph{short-term synaptic
facilitation.}

The synaptic current generated at each synapse then is normally assumed to
be proportional to the fraction of active neurotransmitters, namely $y_{j}(t),$ so the total synaptic current generated in a postsynaptic neuron $i$ is: 
\begin{equation}
I_{i}^{syn}=\sum_{j}^{N}{\cal A}\ y_{j}(t)\ J_{ij}\epsilon_{ij}.
\label{syncur}
\end{equation}
Here ${\cal A}$ is the maximum synaptic current that can be generated at
each synapses, $\epsilon_{ij}$ is the adjacency matrix that accounts for the
connectivity matrix in the neural medium, and $J_{ij}$ are fixed parameters
modulating the synaptic current which can be related, for instance, with
maximal synaptic conductance modifications due to a slow learning process.
In this way, one can choose these \emph{synaptic weights} $J_{ij}$
following, for instance, a Hebbian learning prescription, namely: 
\begin{equation}
J_{ij}=\frac{\kappa}{\langle k\rangle a(1-a)}\sum_{\mu}^{P}(\xi_{i}^{%
\mu}-a)(\xi_{j}^{\mu}-a),\hspace{0.7cm}J_{ij}=J_{ji},\hspace{0.7cm}J_{ii}=0.
\label{pesos}
\end{equation}
Here, $J_{ij}$ contains information from a set of $P$ patterns of neural
activity, namely $\{\xi_{i}^{\mu}=0,1\}$, with $\mu=1,...,P$ and $i=1,...,N$
that are assumed to have been previously stored or memorized by the system
during the learning process. Here $\xi_{i}^{\mu}$ denotes the firing (with
membrane voltage above $V_{th}$) or silent (with $V_{m}$ below $V_{th}$)
state of a given neuron in the pattern $\mu$. The parameter $a$ measures the
excess of firing over silent neurons in these learned patterns, or more
precisely $a=\langle\xi_{i}^{\mu}\rangle_{i,\mu}$. Since $|J_{ij}|$ can be
in general very small and it is multiplying the single synapse currents, we
have considered in (\ref{pesos}) an amplification factor $\kappa=2000$ to ensure a minimum significant effect of the resulting synaptic current (%
\ref{syncur}) in the excitability of the postsynaptic neuron. Moreover, we
also choose a mean node degree factor $\langle k\rangle,$ instead of $N$ in
the denominator of $J_{ij}$ which is more appropriate since it gives a
similar mean synaptic current per neuron for all type of network topologies
considered in this study, including fully connected networks, diluted
networks and complex networks such as scale-free and the classical
Watts-Strogatz small-world networks \cite{WSNature98}.

Following standard techniques from binary attractor neural networks, we can
measure the degree of similarity between a state of the network and a
certain stored activity pattern by means of an $overlap$ function $m^{\mu}(t)$ 
defined as: 
\begin{equation}
m^{\mu}(t)=\frac{1}{aN(1-a)}\sum_{i=1}^{N}(\xi_{i}^{\mu}-a)s_{i}(t),
\label{overl}
\end{equation}
as well as describe the activity of the system through the mean firing rate: 
\begin{equation}
\nu(t)=\frac{1}{N}\sum_{i}s_{i}(t).  \label{mfr}
\end{equation}

In order to visualize if our system is able to respond efficiently to some
input weak stimulus, it is useful to quantify the intensity of the
correlation, during a time window $T,$ between the weak input signal and the
network activity by computing, for instance, the Fourier coefficient at a
given frequency $f$, of the network mean firing rate, that is, 
\begin{equation}
C_{f}=\lim_{T\rightarrow\infty}\frac{1}{T}\int_{0}^{T}\nu(t)e^{ift}dt.
\end{equation}
The relevant correlation, denoted $C(D)$ in the following, it then is
defined as the value of 
\begin{equation}
C(D)\equiv\frac{|C_{f_{s}}|^{2}}{d_{s}^{2}},
\end{equation}
that is, the ration between the power spectrum computed at the frequency of
the input signal $f_{s}$ and the amplitude of this weak signal.

\section*{Results}

\subsection*{The effect of short-term synaptic depression}

As it was stated above, it is important to investigate the mechanisms
involved in the processing of different stimulus by a neural system, in the
presence of noise. This would determine the conditions of ambient or
intrinsic noise at which the transmission of information can be more
efficient mainly, when the relevant information of the stimulus is encoded
in weak signals. This is particularly important in a complex neural system
as it is the brain, where certain brain areas have to respond adequately to
some signals, for instance, arriving from other specific brain areas or the
senses, within a background of noisy activity. Following this aim, we have
first studied how efficient is the processing of noisy weak signals in a
network of $N$ spiking neurons, when it stores a single pattern of neural
activity, and where the synapses among neurons present short-term synaptic
depression. Our study reveals the relevant signals can be processed by the
systems at more than one level of the underlying noise, as it is depicted in
figure \ref{smr}. More precisely, the correlation measure $C(D)$ presents,
two well defined maxima, one at relatively low noise $D_{1}=97.5\ pA$ and
the second at relatively large noise intensity $D_{2}=265\ pA$. Model
parameter values are indicated in the caption of the figure \ref{smr}.

A full description of collective behavior of the network, by means of the
temporal evolution of the mean firing rate and the overlap function compared
with the weak sinusoidal input, and for increasing values of the noise
parameter $D$ along the curve $C(D)$, is depicted in figure \ref{multimo}.
Moreover, raster plots of the network activity, for the same cases shown in
figure \ref{multimo}, are presented in figure \ref{multrp}.

Both figures show (more clearly illustrated in figure \ref{multrp}), that
for relatively low noise the system is able to recall the stored pattern,
which becomes an attractor of the system dynamics. The system, therefore,
shows the associative memory property. When noise intensity is increased to
some given value $D_{1}$ (around $97.5\;pA$ in this figure), the dynamic
regime of the system changes \emph{sharply }to an oscillatory phase where
the network activity periodically switches between a pattern and
anti-pattern configurations. Around this phase-transition point $D_{1},$
these oscillations start to be driven by the weak signal which causes the
first appearing maxima in $C(D)$. This periodically switching behavior
correlated with the weak signal is clearly reflected in the overlap function
and the mean firing rate (see figure \ref{multimo}), and relatively
large-amplitude oscillations in these order parameters with the same
frequency as the sinusoidal weak input signal appear. However, by increasing
further the level of noise $D$, we observe that the correlation with the
input signal is lost. For a further increase of noise around a given value
$D_{2}$ (which in the simulations performed in the figure is about $265\,pA$), 
a second peak in $C(D)$ appears, where a strong correlation of the neural
activity with the input weak signal is recovered. This noise level
corresponds to the critical value of noise at which a second order phase
transition between the oscillatory phase and a disordered phase emerges.

To check the influence of short term depression in the appearance of these
maxima of the correlation function $C(D)$, we have varied the
neurotransmitter recovery time constant $\tau_{rec}$, which is a well know
parameter that permits the tuning of the level of depression at the
synapses. In fact, large recovering time constants are associated to
stronger synaptic depression because the synapses need more time to have
available neurotransmitter vesicles in the ready releasable pool. Therefore,
we repeat the numerical study for several values of the time recovery
constant $\tau_{rec}=250,\ 300,\ 350\ ms$, considering a network of $N=2000$
neurons. The results are depicted in figure \ref{trec}, where $C(D)$ is
shown for different values of $\tau_{rec}$. Two main intriguing effects are
observed. First, the maxima of the $C(D)$, at which there is a high
correlation with the weak signal, appear at lower noise intensities when 
$\tau_{rec}$ is increased. This is due to the fact that, when the level of
synaptic depression is increased, the transitions between ordered and
oscillatory phases and between oscillatory and disordered phases appear at
lower values of noise intensity. This is due to the extra destabilizing
effect over the memory attractors consequence of synaptic depression \cite%
{associ} and to the fact the maxima of $C(D)$ occur precisely at these
transitions points \cite{Gio}. The second effect is that the correlation
with the weak signal (the height of the maxima) increases with the level of
depression, that is, the weak signal is processed with less noise, which is
consequence of the phase transitions points -- and therefore the maxima of 
$C(D)$ -- appear at lower noise values when synaptic depression is increased.

\subsection*{The effect of short-term facilitation}

In general, synapses in the brain and, in particular, in the cortex can
present -- in addition to synaptic depression -- the so called synaptic
facilitation mechanism, that is an enhancement of the postsynaptic response
at short time scales \cite{PNASTSM,NCTsM}. Both \emph{opposite} mechanisms
can interact at the same time scale during the synaptic transmission in a
complex way whose computational implications still are far of being well
understood. The study of the influence of both mechanisms during the
processing of weak signals in a neural medium, constitutes a very suitable
framework to investigate this interplay. With this motivation, we present in
this section a computational study of how synaptic facilitation competing
with synaptic depression influences the detection of weak stimuli in a
network of spiking neurons. In the following computational study, we
consider a fixed time recovery constant $\tau_{rec}=300\ ms$, which is
within the physiological range of the actual value measured in cortical
neurons with depressing synapses \cite{NCTsM}. Also, we take several values
for the characteristic facilitation time constant, namely, $\tau_{fac}=100,\
200,\ 500\ ms, $ and ${\cal U}=0.02.$ The results obtained for the correlation
function $C(D)$ for a network of $N=800$ neurons are
depicted in figure \ref{fac}. In this figure, we can observe a clear
dependence between the level of noise at which maxima in the correlation
function appear and the characteristic facilitation time constant
$\tau_{fac}$. In particular, the figure shows that, as $\tau_{fac}$ increases, the
maxima of $C(D)$ emerge at lower noise intensities. Moreover, one observes
that the intensity of the correlation at its low noise maximum grows
whenever $\tau_{fac}$ is increased.

A possible explanation for this phenomenology is the following: it is well
known, that facilitation favors to reach the stored attractors and their
subsequent destabilization, in auto-associative neural networks \cite{NCJT}.
In other words, synaptic facilitation favors the appearance of the
oscillatory phase. So then, for the same level of noise $D$, more
facilitated synapses induce an easy recovery and posterior destabilization
of the attractors, and therefore, an easy transition to the oscillatory
phase from the memory phase. This, in practice, means that the transition
point between the two phases, appears at lower values of the noise for more
facilitated synapses, and it is precisely at this transition point, where
the low noise maximum of $C(D)$ appears. On the other hand, synaptic
facilitation favors the recovery of the memory attractors with less error 
\cite{NCJT}. In this way, when the transition to the oscillatory phase
occurs, attractors are periodically and transiently recovered with less
error during some time so that, the coherence of the activity of the system
with the weak signals is larger since it is not affected by this extra
source of noise. These findings provide a simple mechanism to control the
processing of relevant information by changing the level of facilitation in
the system which can be done, for instance, controlling the level of calcium
influx into the neuron or by the use of calcium buffers inside the cells.

\subsection*{The effect of network size}

In order to verify the robustness of the results reported above and to
observe the possible effects that may arise due to the finite size of the
system used in our simulations, we have carried out a study of the system
increasing the number of neurons in the network as ${N=400,\ 800,\ 1600,\
2000}$, but maintaining the rest of the parameters and considering spiking
neurons with pure depressing synapses. The computed correlation $C(D)$ for
all these cases is depicted in figure \ref{Ngra}, which reveals that the
main findings of our previous study remain and are independent of the number
of neurons in the system. In fact, different $C(D)$ curves for different
values of $N$ do not present significantly changes neither in their shape
and intensity, nor in the level of noise at which the different maxima of 
$C(D)$ appear. These results permit us to hypothesize that our main findings
here are enough general and could also appear in large populations of
neurons as in cortical slices or even in some brain areas.

\subsection*{The effect of storing many patterns in the network}

In the studies reported in the above sections, we have considered just one
activity pattern of information stored in the maximal synaptic conductances.
We study now, how robust are these findings when the number $P$ of activity
patterns stored in the system increases with all other parameters of the model
unchanged. In our study, we have varied $P$
from 1 to 10 in a network of $N=2000$ neurons with pure depressing
  synapses ($\tau_{rec}=300~ms$) and the corresponding correlation functions $C(D)$ for all
these cases are depicted in figure \ref{pat}. One can appreciate that the
phenomenon remains when $P$ is increased, and while the maximum of $C(D)$
that appears at high noise -- around $D_{2}=265\, pA$ -- does not dramatically
change with $P$, the number of stored patterns has a strong effect on the
maximum of $C(D)$ appearing at low noise. In fact, this maximum appears at
lower level of noise and with more intensity as $P$ is increased.

A possible explanation for this intriguing behavior can be understood as
follows. It is well known that in Hopfield binary neural networks the
increase of the number of stored patterns induces interference among the
memory attractors and therefore constitutes an additional source of noise
that tries to destabilize them \cite{AGS}. This also occurs in our spiking
network and, in the presence of dynamic synapses, this destabilizing effect
results in an early appearance of the transition between the memory phase
and the oscillatory phase and therefore, in the appearance of the first
low-noise maximum of $C(D)$. Then, at relatively low levels of ambient noise 
$D$ the transition from below to the oscillatory phase will occur at lower
values of $D$ in a network that stores a larger number of patterns. At
relatively large values of the ambient noise, however, the main
destabilizing effect is due to the ambient underlying noise, and therefore,
the effect of increasing $P$, although is also present, it is less
determinant. On the other hand, the amplitude of the maximum of $C(D)$
increases with $P$ because, as explained above, the transition among memory
and oscillatory phases occurs at low value of the ambient noise for $P$
larger. Then, the thermal fluctuations in the memory phase and during the
phase transition to the oscillatory phase are lower. In this way, during the
oscillations starting at the transition point, the attractors are recovered
transiently with less error which induces the coherence with the weak signal
to be larger.

\subsection*{The effect of the asymmetry of the stored pattern}

The features of the pattern-antipattern oscillations which characterize the
oscillatory phase in neural networks with dynamic synapses (including
short-term facilitation and depression), are highly dependent on the
particular symmetry in the number of active and silent neurons in the stored
pattern. This is controlled by the parameter $a$ (introduced in the
definition of the synaptic weights (\ref{pesos})). In all the results
reported in previous sections, we have considered $a=0.5$, which causes an
oscillatory phase characterized by a regime of symmetric oscillations
between an activity state correlated with the stored pattern and another
with the same level of activity correlated with the antipattern. If we
consider $a\neq0.5$, an asymmetry will be induced in the mean network
activity, that is, there will be an excess of 1's over 0's or vice versa
during pattern-antipattern oscillations. In fact, oscillations occurs
between an high activity (Up) state and a low activity (Down) state.
Moreover, this asymmetry in the activity of the stored pattern can have a
strong influence in the phase diagram of the system and can cause even that
the oscillatory phase does not emerge. Since the phenomenology of interest
here -- namely the emergence of an network activity correlated with a weak
subthreshold stimulus -- is highly dependent on the transition points at
which the system moves over different phases, it is reasonable to think that
the parameter $a$ will have a strong influence on it.

We have performed a computational study in a network of $N=800$ neurons with
pure depressing synapses ($\tau_{rec}=300~ms$) to investigate this particularly
interesting issue, for which we have considered a single stored pattern $P=1$
with $a=0.40,\ 0.42,\ 0.43,\ 0.45,\ 0.47,\ 0.48$ and whose results are
depicted in figure \ref{a}. We observe here a very interesting and
intriguing effect in the shape of the correlation $C(D)$ when $a$ is varied.
The maximum in the correlation between the network activity and the weak
signal at low noise tends to disappear as the value of $a$ decreases from
the symmetric value $a=0.5$. In fact, when $a<0.45$ the correlation $C(D)$
drops abruptly at that point, around $D_{1}=100\, pA$. As we have explained
above, a large level of asymmetry in the stored pattern could impede the
appearance of the Up/Down transitions characteristic of the oscillatory
phase which, therefore is absent. The consequence is that, there is not a
transition point between a memory phase and an oscillatory phase which
impedes the emergence of the low noise maximum of $C(D)$. On the contrary,
the second maximum of $C(D)$, which appears at high noise around 
$D_{2}=265\, pA$, remains invariant for all values of $a$ studied here. The
explanation to this second situation is also simple because, although
asymmetry in the stored pattern is present, the phase diagram of the system
still presents a phase of memory retrieval at low noise and a non-memory
phase at large noise, separated by a second order phase transition point
around which the second maximum of $C(D)$ is originated.

\subsection*{The effects of the underlying network topology}

In previous sections, we have considered for simplicity -- as our system
under study -- a fully connected network of spiking neurons. This is far to
be the situation in actual neural systems, where neurons are not all
connected to each others. In fact, biological neural systems are
characterized by a underlying complex network topology which is consequence
of different biophysical processes during their developing, including among
others, exponential growth at early stages of developing and posterior
synaptic pruning processes \cite{John}. All these processes are also
influenced by limitations in energy consumption in the system. In this
section, we explore if the emergence of several maxima, as a function of the
underlying noise, in the correlation between the network activity and some
weak subthreshold stimulus, is altered when more realistic network
topologies are considered.

We have considered first the case of a random diluted network. We can
configure this network topology starting, for instance, starting with a
fully connected network and then removing randomly a certain fraction $\delta
$ of the synaptic connections. In figure \ref{dil}A, it is depicted the
resulting correlation function $C(D)$ for single realizations of diluted
networks generated in this way with $N=800$ and $\delta=$10\%, 20\%, 30\% and 40\%. The figure
illustrates two main findings. First, the robustness of the main emergent phenomena described in
the above sections also in this type of diluted networks, and second that,
as the dilution grows and a higher fraction of connections is removed, both
maxima of $C(D)$ appear respectively at lower levels of noise. Moreover, if
dilution is too high, it seems that the maximum appearing at low noise tends
to disappear. This only can be consequence that the stable memory attractors
lose stability -- due to strong dilution -- and disappear in the presence of
ambient noise, in such a way that only an oscillatory phase and non-memory
phases are present.

In our analysis with a diluted network, we performed dilution starting with
a fully connected network where $\langle k\rangle=N.$ To avoid the possible
effect of a factor $1/N$ normalizing the synaptic weights (\ref{pesos})
during dilution, we have done an additional analysis considering a diluted
network with a normalizing factor in the weights $\langle
k\rangle=(1-\delta)^{2}N,$ which is the mean connectivity degree in the
resulting diluted network with $\delta$ being the probability of a link to
be removed during the dilution process. The corresponding results are
summarized in the figure \ref{dil}B. One can observe also that results are
similar for this second type of dilution, that is, the low noise maximum of 
$C(D)$ moves toward lower values of the ambient noise and even can disappear
as dilution is increased. The main difference with the first type of
dilutions is, however, that the level of noise at which the high noise
maximum of $C(D)$ is not dramatically affected by dilution. In figure
\ref{dil}B, the correlation curves $C(D)$ have been obtained after averaging
over $10$ realizations of a network of $N=200$ neurons with pure depressing
synapses (with $\tau_{rec}=200~ms$). 

Diluted networks, however, are homogeneous and do not introduce complex
features which could induce more intriguing behavior in the system. Even
more interesting and realistic concerns the case of networks with complex
topology such as the so called scale-free networks, where the node degree
probability distribution is $p(k)\sim k^{-\gamma}$, with $k$ being the node
degree. In fact, it has been recently reported that these complex topologies
can induce additional correlation with the network activity and the
processed weak stimulus due to the network structural heterogeneity \cite%
{kra}. Figure \ref{sfn} summarizes our main results concerning the case of
complex networks with scale-free topology. Correlation curves $C(D)$ have been
obtained after averaging over $10$ realizations of a scale-free network of
$N=200$ neurons with pure depressing synapses (with $\tau_{rec}=200~ms$) and all
other model parameters as in figure \ref{smr}. 
We see that similarly to the
cases studied above, two maxima in the correlation function $C(D)$ emerge -
for two well defined values of the underlying noise -- also in scale-free
networks. However, we do not observe the emergence of an additional maximum
of $C(D)$ which could be induced only by the topology. As it has been
reported in \cite{kra}, this maximum should appear at low values of the
ambient noise, for $\gamma\sim3$. The existence of a robust oscillatory
phase when dynamic synapses are considered and a phase transition between
memory and this oscillatory phase at relatively low values of the ambient
noise could hidden the appearance of the this maximum due to the existence
of a low noise maximum of $C(D)$ around this phase transition. In any case,
as it is depicted in figure \ref{sfn}, the emergence of two maxima in $C(D)$
is a robust phenomenon also in these complex scale-free network topologies
for a wide range of the relevant network parameters such as the exponent of
the network degree distribution (figure \ref{sfn}A) and the mean
connectivity in the network (figure \ref{sfn}B). Interestingly is that the
low noise maximum seem to start to emerge for values of $\gamma\gtrsim3$ and
values of the mean connectivity $\langle k\rangle\gtrsim15-20.$ These values
corresponds to realistic ones, since, for instance, most of the actual
complex networks in nature have degree distributions with $\gamma$ between 2
and 3 \cite{Newmanreview}. Moreover neurons in the brain of mammals have
large connectivity degrees and realistic values of the mean structural
connectivity in cortical areas of mammals has been reported to be around 
$\langle k\rangle\approx20$ \cite{sporns07}.

Finally, we have consider in our study the case of a complex network with
the small-world property. A prominent example of such type of networks is
the so called Watts-Strogatz (WS) network \cite{WSNature98}. These networks
are generated starting with a regular network where each node, normally
placed in a circle, has $k_{0}$ neighbors. Then, with some probability
$p_{r}$, know as probability of rewiring, each link among nodes in this regular
configuration is rewired to a randomly chosen node in the network avoiding
self-connections and multiple links among two given nodes. In this way for 
$p_{r}=0,$ one has a regular network with $p(k)=\delta(k-k_{0})$ and for 
$p_{r}=1$ one has a totally random network with $p(k)$ being a Gaussian
distribution centered around $k_{0}.$ Note that for varying $p_{r}$ one
always has $\langle k\rangle=k_{0}.$ We have placed neurons defined by the
dynamics (\ref{IF}) in such WS networks and studied as a function of the
underlying noise the emergence of correlations between the network activity
and some subthreshold weak signals by means $C(D)$. The results are
summarized in figure \ref{sw_network} where each $C(D)$ curve has been
obtained after averaging over $10$ realizations of a network with $N=200$
neurons and pure depressing synapses with $\tau_{rec}=100~ms.$ One can see that the appearance
of several maxima for $C(D)$ occurs for values of $p_{r}\gtrsim0.5.$ More
precisely, the low noise maximum does not emerge for low rewiring
probabilities, which clearly indicates that the memory phase does not appear
for such small values of $p_{r}$ for the whole range of noise $D$ considered
here. Also this finding suggests the positive role of long range connections
- that only can emerge with high probability when $p_{r}$ is high -- for the
existence of such low noise maximum in $C(D)$. In fact, the emergence of a
memory phase can be only understood when in the network appear such long
range connections since the stored memory patterns involve these type of
spatial correlations among active and inactive neurons.

\subsection*{Use of more realistic weak signals\label{poisson_signals}}

In actual neural systems is expected that relevant signals arrive to a
particular neuron in the form of a spike train, with relevant information
probably encoded in the timing among the spikes. In this sense, the use of a
sinusoidal weak current to explore how the system detect it in all cases
considered above could not be the most realistic assumption (it could be
enough realistic if relevant information will be encoded in subthreshold
oscillations instead that in the precise timing of the spikes). To
investigate the ability of the system to detect and process the with more
realistic weak signals in the presence of noise we have considered the input
weak signal as an inhomogeneous Poisson spike train with mean firing rate 
$\lambda(t)=\lambda_{0}[1+a\mbox{sin}(2\pi f_{s}t)],$ being $\lambda_{0},a$
positive constants. In this way, relevant information is encoded as a
sinusoidal modulation of the arrival times of the spikes in the train.
Figure \ref{figpois}A depicts the coherence among mean firing rate in the
network and this weak signal (which is shown in the top graph of figure 
\ref{figpois}B). The correlation curve $C(D)$ has been obtained after
averaging over $20$ realizations of a fully connected network with $N=400$ 
neurons and pure depressing synapses with $\tau_{rec}=300~ms.$ The figure clearly illustrates that also in this more realistic
case the system present a strong correlation with the weak signal at
different levels of noise at which phase transitions among different
non-equilibrium phases appear (see time series for increasing level of noise
from top to bottom in figure \ref{figpois}B). These are a memory phase where
active neurons in the stored memory pattern are strongly synchronized 
($D=60\, pA$) (population burst regime), a transition point characterized by
signal driven high-activity (Up state)/low-activity (Down state)
oscillations ($D_{1}=85.4\, pA$), a phase of intrinsic Up/down oscillations 
($D=160\, pA$), a critical point toward a non-memory phase characterized by
signal driven fluctuations plus thermal fluctuations (at
$D_{c}=D_{2}=261\,pA$), and a non-memory phase, or asynchronous phase, 
characterized by constant firing rate with Gaussian thermal fluctuations 
(for instance at $D=500\, pA$). In fact, in figure \ref{figpois}C it is 
depicted the difference in steady state features of the behavior for the two 
last cases. That is, at the critical point $D_{2}$ the stationary distribution 
of the resulting mean firing rate has a bias toward positive fluctuations 
induced at the exact arrival time of the weak signal spikes. This as evidenced 
by the disagreement between this distribution (red curve in figure 
\ref{figpois}C) and the shaded red area, which represents the best fit to a 
Gaussian distribution. On the other hand, during the asynchronous state for 
$D=500\,pA$ the same steady state distribution (green line in figure 
\ref{figpois}C) is clearly Gaussian (shaded blue area) without presenting a 
bias with the timing of signal spikes.

\section*{Discussion}

We investigated in great detail by computer simulations the processing of
weak subthreshold stimuli in a auto-associative network of spiking neurons
- $N$ integrate and fire neurons connected to each others by dynamic
synapses using different network configurations -- competing with a
background of ambient noise. In particular, we studied the role of
short-term synaptic depression in the efficient detection of weak periodic
signals by the system as a function of the noise. Our results show the
appearance of several well defined levels of noise at which there is a
strong correlation between the mean activity in the network and the weak
signal. More precisely, in the range of noise intensities considered in this
study, the transmission of the information encoded in the weak input signal
to the network activity is maximum when noise intensity reaches two certain
values.

The maximum or peak appearing at relatively low levels of ambient noise 
$D_{1}$, corresponds to a transition point where the activity of the network
switches from a memory phase -- in which a stored memory pattern is retrieved
- to an oscillatory phase where the system alternatively is recalling the
stored pattern and its anti-pattern in a given aperiodic sequence. Thus, at
this level of noise and in the presence of the weak signal this oscillatory
behavior of the network activity becomes correlated with the signal
oscillating at its characteristic frequency. This fact provides an efficient
mechanism for processing of relevant information encoded in weak stimuli in
the system since at this transition point, for instance, the system could
efficiently recall different sequences of patterns of information according
to predefined input signals.

On the other hand, the second maximum which appears at relatively high level
of ambient noise, namely $D_{2}$, emerges around a second order phase
transition between the oscillatory phase explained above and a disordered or
non-memory phase, where the system is not able to recall any information
stored in the patterns. Although the resulting network activity around this
maximum is highly noisy, it is strongly correlated with the weak signal,
appearing a modulation of the noisy activity that follows the signal
features (see the case $D_{2}=265\, pA$ in figure \ref{multimo}).

We also studied in detail, the influence that the particular level of
synaptic depression could have in the appearance of the two maxima in the
correlation between the network activity and the weak stimulus. By changing
the recovery time constant of the active neurotransmitters, for instance, we
observe that the longer the synapses take to recover the neurotransmitters
(larger $\tau _{rec}$), the lower level of ambient noise is needed to induce
to reach the different maxima of the corresponding correlation function (see
the figure \ref{trec}).

Furthermore, we have observed that short-term facilitation competing with
short-term depression at the synapses induces also additional intriguing
effects in the way system responds to the weak stimulus in the presence of
noise. Our results reveal, that for larger values of the characteristic time
constant characterizing the facilitation mechanism, namely $\tau_{fac}$, the
maxima in the correlation $C(D)$ between the network activity and the weak
signal appear at lower levels of ambient noise than in the case that only
synaptic depression is considered. In addition, we can observe that the
correlation at the low noise maximum amplifies when the level of synaptic
facilitation increases (see figure \ref{fac}). Both phenomena can be
understood taking into account that facilitation favors the retrieval of
information in the attractors and their posterior destabilization, which is
the origin of the oscillatory phase. Then, an increase of facilitation moves
the transition point between the memory phase and the oscillatory phase
towards lower values of the ambient noise $D$. Also, facilitation favors the
recovery of the memory attractors with less error, which implies that, when
the coherence with the weak signals emerges in the network activity, it is
affected by less sources of noise, and therefore, it increases for large
facilitation.

We proof as well the robustness of the results reported here, by checking
that the appearance of several maxima in the correlation $C(D)$ between the
network activity and weak subthreshold stimuli, around some phase
transitions points remains for larger number neurons $N$ or number of stored
patterns $P$. Our study reveals that our main findings are independent of
the network size (see figure \ref{Ngra}) and therefore could probably also
obtained in actual neural media. Nevertheless, when we study the correlation 
$C(D)$ for different number of stored patterns, some new effects appear. In
fact, the low noise maximum of $C(D)$ tends to disappear when the number of
stored patterns is increased. The reason is, that an increase in $P$ induces
the appearance of the oscillatory phase at lower values of the ambient noise 
$D$ and, consequently, the development of this maximum of $C(D)$ occurs at
lower noise (see figure \ref{pat}). On the other hand, the appearance of the
second high noise maximum of $C(D)$ is not significantly affected by an
increase of $P$ (see also figure \ref{pat}).

Another interesting result is obtained when we consider an asymmetric stored
memory pattern, that is, $a\neq0.5$, so that there is an excess of 1's over
0's in the stored pattern or vice versa. In figure \ref{a}, we can see that
by increasing the pattern asymmetry (reducing $a$), the low noise maximum of 
$C(D)$ decreases. This is mainly due to the fact that oscillations between
the high-activity and low activity states, characterizing the oscillatory
phase, tend to be less visible or even disappear -- the network activity is
quasi clamped in the memory attractor -- for more asymmetric stored patterns.
In fact, if we consider $a<0.45$, the low noise maximum of $C(D)$ drops
abruptly and the system is not able anymore to process the information
encoded in the weak signal at this level of ambient noise (see figure \ref{a}).

We have performed also a complete study of how our main findings are
influenced by the given network topology. As a first step in this research
line, we have considered the case of diluted networks. We built different
types of diluted networks by erasing a fraction of links at random in a
fully-connected network. In all cases that we have considered, the main
results still emerge, that is the existence of several maxima in the
correlation $C(D)$ between the network activity and the weak stimulus at
some precise levels of noise around non-equilibrium phase transitions. If
the fraction of erased synaptic connections is larger, these maxima of $C(C)$
appear at lower level of ambient noise. Moreover, there is a tendency that
leads to the complete disappearance of the low noise maximum for strong
dilution (see figure \ref{dil}). These findings are due to the fact that,
dilution of synaptic links diminishes the memorization and recall abilities
of the network (since memory pattern features are stored in these links).
The consequence is that, in more diluted networks, the memory phase appears
at lower values of ambient noise, and can even disappear in absence of noise
for very diluted networks. Secondly, we have consider also the case of
complex networks with scale-free properties in the degree distribution and
with the small-world property. In all cases there is a wide range of the
relevant networks parameters, such as the exponent of the scale-free degree
distribution, the mean connectivity in the network and the rewiring
probability in the the case of the WS small world network, for which $C(D)$
shows also similar maxima at given noise levels where the system efficiently
process the weak stimulus. Moreover this range of parameters is consistent
with those measured in actual neural systems.

Note that the consideration here of some complex networks with scale-free
topology, introduces node-degree heterogeneity in the system. This fact
induces also neuron heterogeneity in the sense that more connected neurons
can be more excitable than less connected neurons, so one can have different
levels of neuron excitability in the system. However, we have seen that even
in this case there is a wide range of model parameters where the relevant
phenomenology here still emerges. This tell us that in a more general
scenario where different types of neuron heterogeneity are considered, the
phenomena reported here also will emerge.

After our analysis in the present work, we conclude that the efficient
detection or processing of weak subthreshold stimuli by a neural system in
the presence of noise can occur at different levels of this noise intensity.
This fact seems to be related to the existence of phase transitions in the
system precisely at this levels of noise, a suspicion which is presently
been analyzed in greater detail. In the cases studied here, within the range
of noise considered, there is a maximum in the correlation $C(D)$ of the
network activity with the stimulus which seems to correspond to a
discontinuous phase transition (the maximum at relatively low ambient
noise) as well as another maximum appearing around a continuous phase
transition (the maximum at relatively high noise). The difference in the
type of the emerging phase transition determines the way the weak
subthreshold stimulus is processed by the neural medium. 

We hope to study
next the computational implications that each one of theses maxima induces
and their possible relation with high level brain functions. In particular,
in some preliminary simulations reported in the present work with more
realistic Poissonian signals, a detailed inspection of the network activity
temporal behavior, compared with the weak signal time series around the low
noise maximum, show a strong resemblance with working memory tasks, where
relevant information encoded in the input signal is maintained in the
network activity during some time, even when the input signal has
disappeared. Also in the case of several memory pattern stored in the system
and around this low noise maximum, the system could process a given sequence
of patterns encoded in the stimulus. Precisely at this level of noise, a
non-equilibrium phase emerges characterized by continuous sequence of jumps
of the network activity between different memories, which could be
correlated with a particular sequence of memories in the presence of an
appropriate stimulus. On the other hand, at the second resonance peak is the
precise timing between input spikes which are detected and processed by the
network activity.

Finally, we mention that it would be interesting to investigate if the
relevant phenomenology reported in this work could emerge naturally in
actual systems. In fact, recent data from a psycho-technical experiment in
the human brain \cite{jap} can be better interpreted, using different
theoretical approaches and dynamic synapses, considering the existence of
several levels of noise at which relevant information can be processed 
\cite{torresNJP,Gio}. In figure \ref{data_exp} it is shown how these 
experimental data can be also interpreted in terms of the correlation function 
$C(D)$ obtained within the more realistic model approach reported in this paper,
that is, a complex network of spiking neurons. This should serve as
motivation to study in depth how neural systems process weak subthreshold
stimuli in a more biological and realistic scenario. For instance, one could
consider conductance based neuron models instead of the simplified integrate
and fire model used here or conceive more realistic stimuli, and other
complex network topologies. The last could include, for instance, different
type of node degree-degree correlations 
\cite{johnsonprl,sebastpre11,lucioli14} or network-network correlations 
constituting a \emph{multiplex} structure as a recent work suggests to occur
in the brain \cite{reisnature14}. All these additional considerations could
provide some new insights in order to design a possible experiment easily
reproducible by biologists to investigate the emergence the phenomena
reported here in actual neural systems. On the other hand, the relation of
the relevant phenomenology reported in this study with the existence of
different phase transitions in our system could be of interest for
neuroscientists to investigate the existence of phase transitions in the
brain.

\section*{Acknowledgments}

We acknowledge useful comments by Matjaz Perc and Claudio Mirasso, and 
financial support from the Spanish MINECO project FIS2013-43201-P.


\newpage

\section*{Figure Legends}

\begin{figure}[h!]
\caption{Correlation function $C(D)$ for a single
realization of a fully connected network of $N=1600$ spiking neurons with
pure depressing synapses with $\tau_{rec}=300\ ms$ and one stored
pattern $P=1$. Other parameters were $a=0.5$, $\kappa=2000,$ 
$I_{0}=100\ pA$, $d_{s}=20\ pA$, $f_{s}=1.5\ Hz,$ ${\cal U}=0.5$, 
${\cal A}=45\ pA$, $\tau_{in}=3\ ms$.}
\label{smr} 
\end{figure}

\begin{figure}[h!]
\caption{Temporal behavior of the mean firing rate and
the overlap function for five different values of the noise intensity $D$
along the curve $C(D)$ depicted in figure \ref{smr}. This
shows a tendency to the coherence between network activity and the weak
stimulus around two values of the noisy intensity, namely $D_{1}=97.5\ pA$
and $D_{2}=265\ pA$. The weak sinusoidal stimulus is also shown on the top
panels.}
\label{multimo}
\end{figure}

\begin{figure}[h!]
\caption{Raster plots of the network activity for different values of the
noise intensity $D$ for the network realization with the correlation 
$C(D)$ depicted in figure \ref{smr}. The stored pattern
in such that for $i$=0,...,799 $s_{i}=1$, that is neurons are active, and
for $i$=800,...,1600 $s_{i}=0$ that is, these neurons are silent in the
pattern. At the first maximum of $C(D)$ that occurs around $D_{1}=97.5\, pA$
there are oscillations of the network activity around the stored pattern and
its anti-pattern which are correlated with the weak stimulus. At a second
critical level noise $D_{2}=265\,pA$ a transition between the above
oscillatory phase and a disordered phase appears, and a second maximum of 
$C(D)$ emerges where a very noisy mean activity in the network also is
correlated with the weak stimulus. All depicted panels corresponds to the
same cases shown in figure \ref{multimo}.}
\label{multrp}
\end{figure}

\begin{figure}[h!]
\caption{The behavior of the correlation $C(D)$ for a single
realization of a network of $N=2000$ neurons with different levels of
synaptic depression at the synapses monitored by varying $\tau_{rec}$. 
The figure shows that an increase of $\tau_{rec}$ causes the
maxima of $C(D)$ to appear at lower noise intensities. Other parameters were
as in figure \ref{smr}.}
\label{trec}
\end{figure}

\begin{figure}[h!]
\caption{The correlation $C(D)$ obtained for a single realization of
the system with $N=800$ neurons considering both short-term depression and
short-term facilitation processes at the synapses with ${\cal U}=0.02.$ The
figure illustrates that both maxima of $C(D)$ appear at lower noise
intensities as well as that the amplitude of the first maximum grows when 
$\tau_{fac}$ is increased. Other parameters were as in figure \ref{smr}.}
\label{fac}
\end{figure}

\begin{figure}[h!]
\caption{The correlation $C(D)$ for different number of neurons in the
network. The figure shows the same maxima and shape for $C(D)$ independently
of $N$ which confirms that the results obtained are independent from the
size of the network considered. Each curve has been obtained with a
single realization of the corresponding network. Other parameters were as
in figure \ref{smr}.}
\label{Ngra}
\end{figure}

\begin{figure}[h!]
\caption{Correlation function $C(D)$ for different number of stored patterns 
$P$ in a network with $N=2000$ neurons. This shows that as $P$ increases the
low-noise maximum of the correlation $C(D)$ increases in intensity and
appears at lower level of noise while the second correlation maximum remains
unchanged. Each curve has been obtained with a single realization of
the corresponding network. Other parameters values were as in figure
\ref{smr}.}
\label{pat}
\end{figure}

\begin{figure}[h!]
\caption{Changes in the shape of the correlation $C(D)$ for a network with 
$N=800$ neurons considering different levels of asymmetry in the network
activity encoded in the stored pattern as measured with the parameter $a$
(see main text for the explanation). Panel A depicts correlation curves for 
$a\leq0.45$, showing that the correlation $C(D)$ drops around $D=100\,pA$ and
another maximum starts to emerge at $D=65\,pA$. Panel B illustrates $C(D)$
curves for $a\geq 0.45$, showing that the low noise maximum at $D=100\, pA$
looses intensity as $a$ decreases. Each curve has been obtained with
a single realization of the corresponding network. Other parameters were as
in figure \ref{smr}.}
\label{a}
\end{figure}

\begin{figure}[h!]
\caption{Behavior of the correlation $C(D)$ for a diluted
network. (A) As the fraction $\delta$ of removed connections
increases both maxima of $C(D)$ appear at lower noise intensity. The network
size was set to $N=800$ and a factor $N$ instead of $\langle k\rangle$ is
considered in the definition (\ref{pesos}). Each curve has
been obtained with a single realization of the corresponding network. Other
parameters were as in figure \ref{smr}. (B) In this case simulations
has been performed considering diluted networks using the same procedure
than in panel A, but with less level of synaptic depression, $\tau_{rec}=200\;
ms,$ and synaptic weights normalized with a factor $\langle k\rangle=(1-\delta)^{2}N.$ Also in this case, the corresponding $C(D)
$ curves has been obtained for $N=200$ and averaging over $10$ different
networks.}
\label{dil}
\end{figure}

\begin{figure}[h!]
\caption{Detection of weak subthreshold stimulus as a
function of noise $D$ in spiking networks with complex scale-free
topologies. (A) Correlation curves $C(D)$ obtained for different values of
the exponent $\gamma$ of the degree connectivity distribution 
$p(k)\sim k^{-\gamma}$ in a network with $\langle k\rangle=20.$ The
figure shows that around $\gamma=3$ multiple maxima in $C(D)$ start
to emerge. (B) The same correlation curves $C(D)$ for scale free networks
with $p(k)\sim k^{-3}$ and different values of the mean connectivity degree 
$\langle k\rangle.$ Again for more realistic large values of $\langle k\rangle
$ around 20 multiple maxima in $C(D)$ start to appear. All curves have been
obtained for $\tau_{rec}=200\ ms$ in a network with $N=200$ neurons
and after averaging over 10 different network realizations. Other
parameters as in figure \ref{smr}.}
\label{sfn}
\end{figure}

\begin{figure}[h!]
\caption{Detection of weak subthreshold stimulus as a
function of noise $D$ in a WS small-world network a function of the rewiring
probability $p_r.$ The network has $N=200$ neurons each one with 
$k_{0}=\langle k\rangle=20$ neighbors and depressing synapses with
$\tau_{rec}=100~ms$. Different correlation functions $C(D)$ have been obtained
after averaging over 10 different network realizations. Other parameters as in figure \ref{smr}.}
\label{sw_network}
\end{figure}

\begin{figure}[h!]
\caption{Detection of more realistic Poissonian weak
stimuli as a function of the noise noise $D.$ (A) Corresponding correlation 
$C(D)$ for a network of $N=400$ neurons depicting several maxima similarly to
the case of sinusoidal stimuli. The curve has been obtained after averaging
over $20$ network realizations. The weak signal current $I_i^{ext}$
is the resulting of an inhomogeneous Poissonian train of spikes of amplitude 
$d_s=10\,pA$ injected in each neuron (see panel B top graph) at frequency 
$\lambda(t)=\lambda_{0}[1+a\mbox{sin}(2\pi f_{s}t)]$
with $\lambda_{0}=70\,Hz$ and $a=0.75.$ (B) Time-series of the
network mean firing rate for increasing values of the noise intensity $D$
along the correlation curve depicted in panel A. (C) Normalized histograms
of the network mean-firing rate at the high noise maximum of $C(D)$ (red
curve) corresponding to the critical point $D_{c}=261\,pA$ and for a higher
level of noise, namely $D=500\,pA$, above this critical point (green curve).
The colored areas corresponds to the best Gaussian fitting of both
histograms. Other parameters were as in figure \ref{smr}.}
\label{figpois}
\end{figure}

\begin{figure}[h!]
\caption{Fitting experimental data in the auditory cortex
to our model. The experimental data (symbols with the corresponding error
bars) reported in \cite{jap} are compared as a function of noise
with the correlation function $C(D)$ described in figure \ref{smr} 
corresponding to a single realization of a spiking network of 
$N=2000$ integrate and fire neurons (red solid line) and dynamic synapses.
The experimental data $C$ (in arbitrary units) has been multiplied by a
factor $10^{4},$ and the noise amplitude $M$ (in $dB$) has been transformed
in our noise parameter $D$ using the nonlinear relationship $D=D_{0}+\frac{M\eta}{2}\left\{ 1+\mbox{erf}\left[(M-M_{0})/\sqrt{2}\sigma_{M}\right]\right\} $ with $D_{0}=0.1\,pA,\eta=0.71\,(pA/dB),\, M_{0}=50dB$ and $\sigma_{M}=140\, dB.$}
\label{data_exp}
\end{figure}

 \vspace*{15cm}

\newpage
\vspace*{15cm}

\begin{center}

\includegraphics{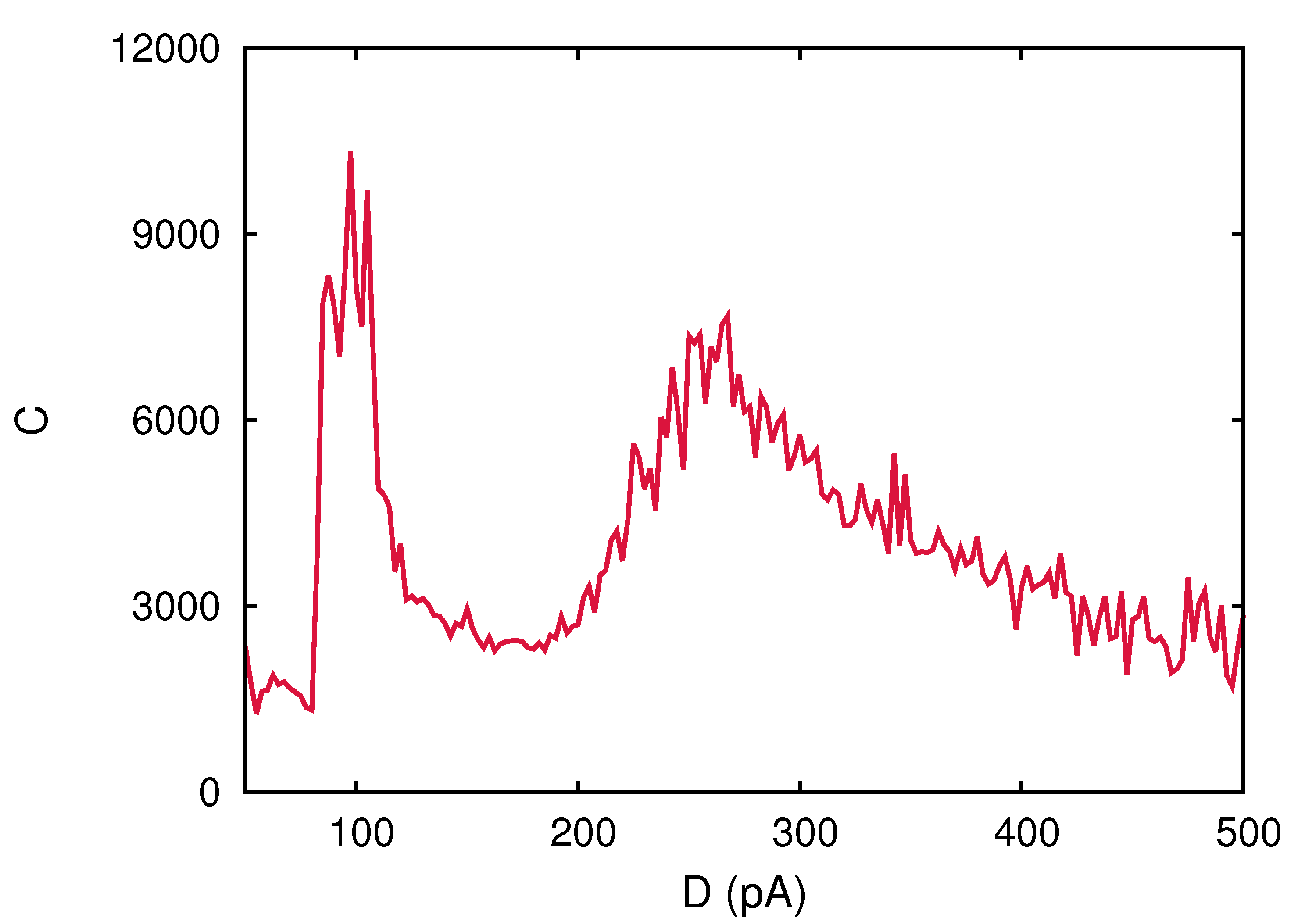}

{\Large Figure 1} 
\end{center}

\newpage

\begin{center}
\includegraphics[scale=0.9]{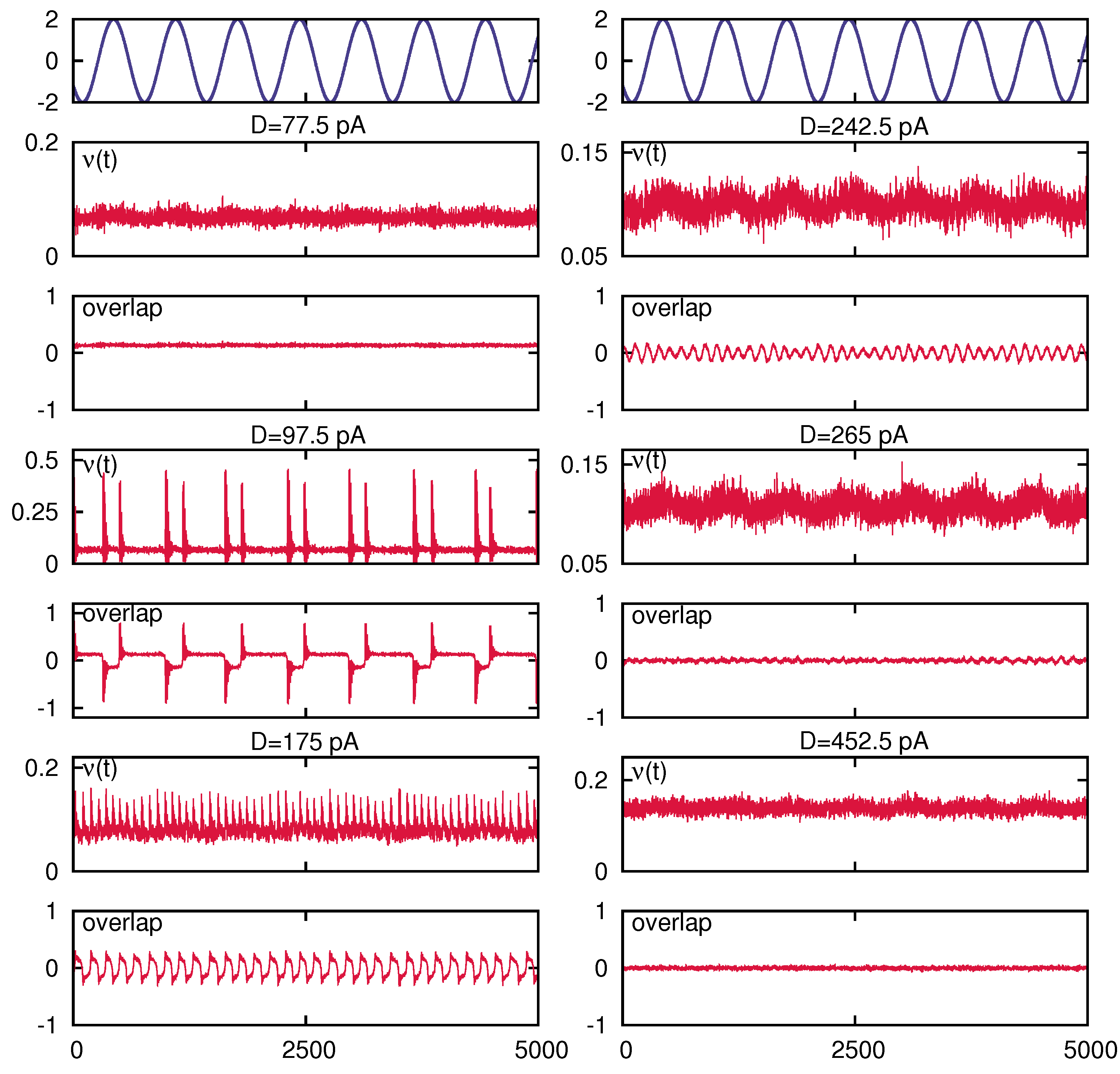} 

{\Large Figure 2} 
\end{center}

\newpage

\begin{center}
\includegraphics[scale=0.9]{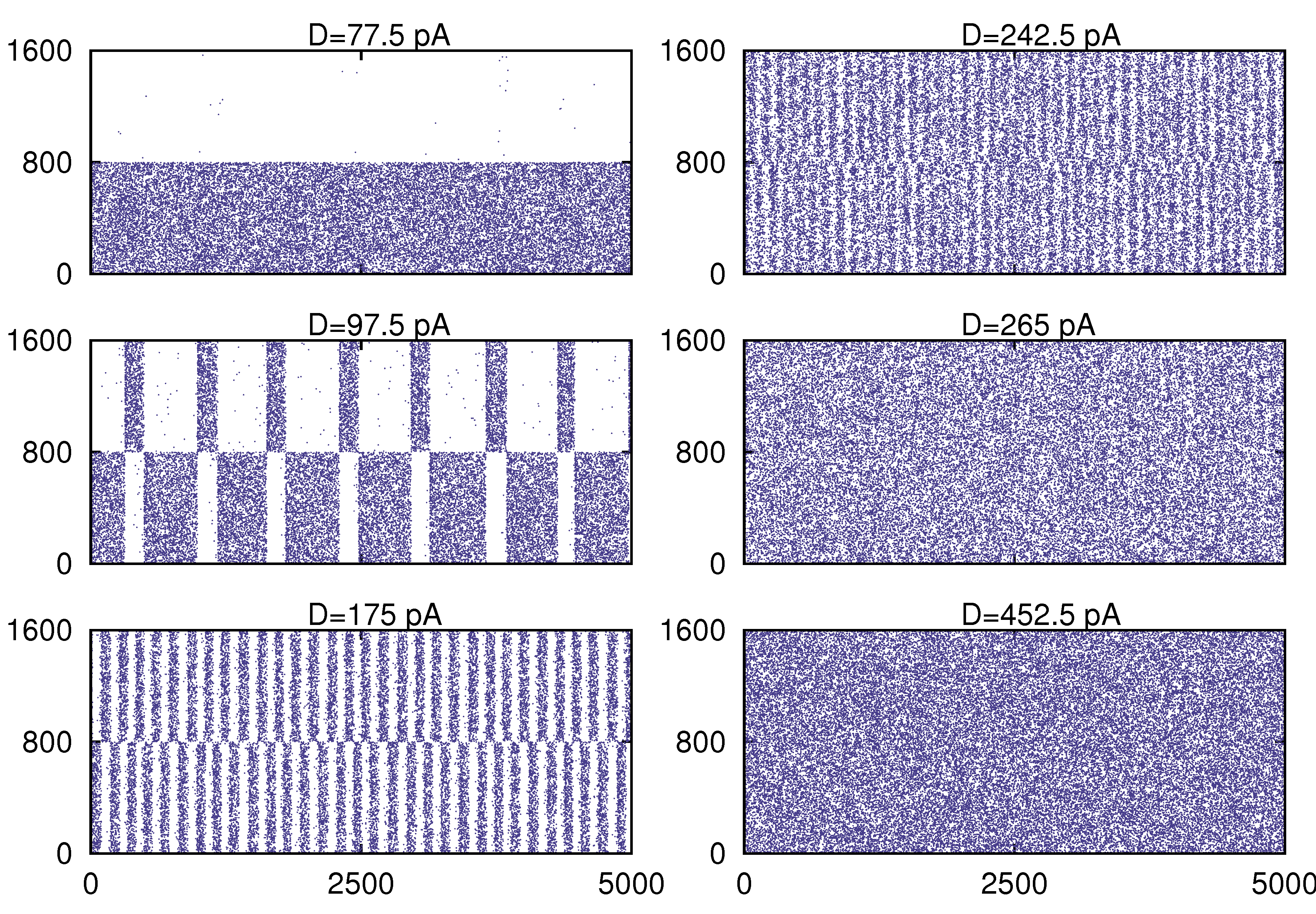} 

{\Large Figure 3} 
\end{center}

\newpage

\begin{center}
\includegraphics{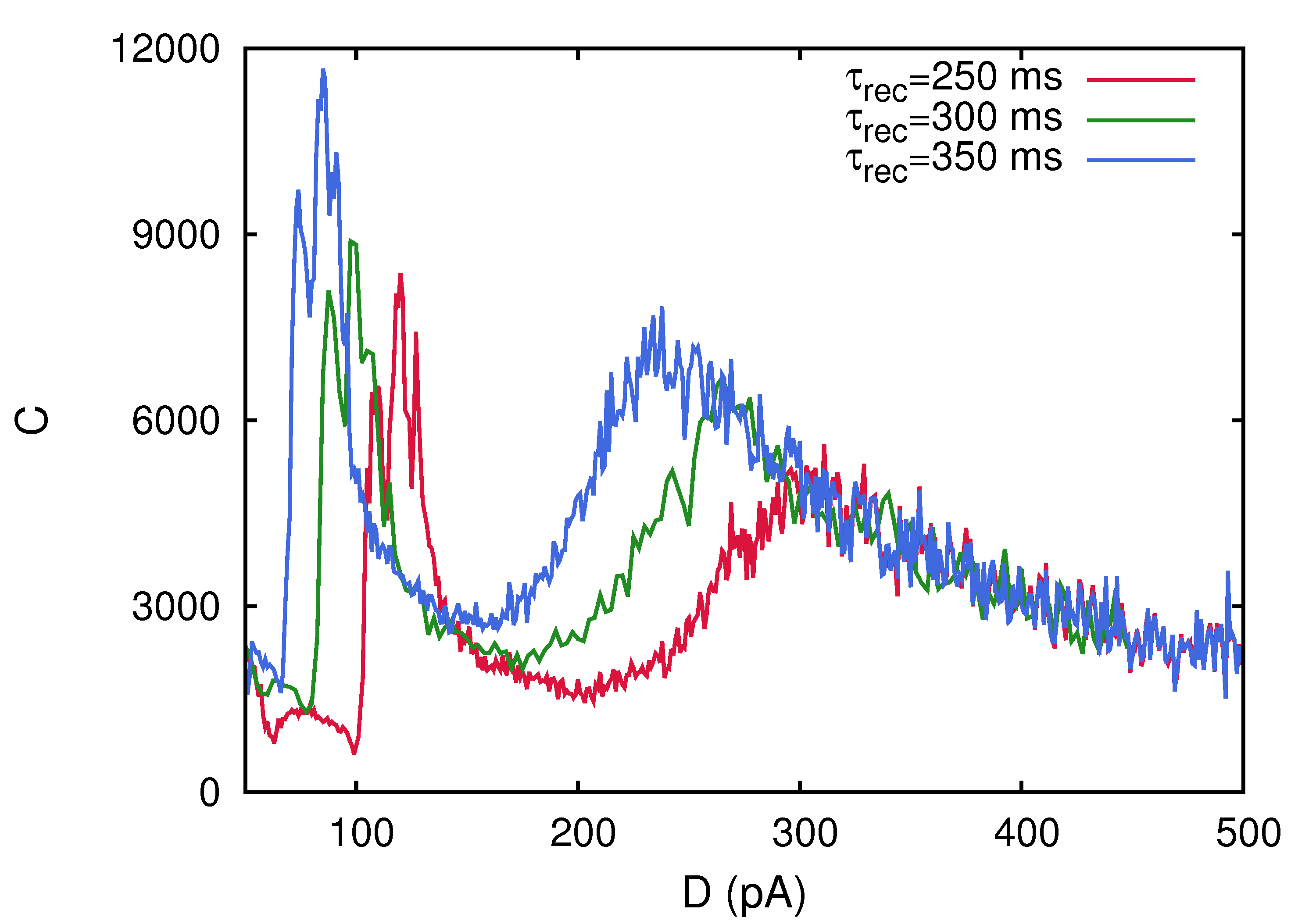} 

{\Large Figure 4} 

\end{center}

\newpage

\begin{center}
\includegraphics{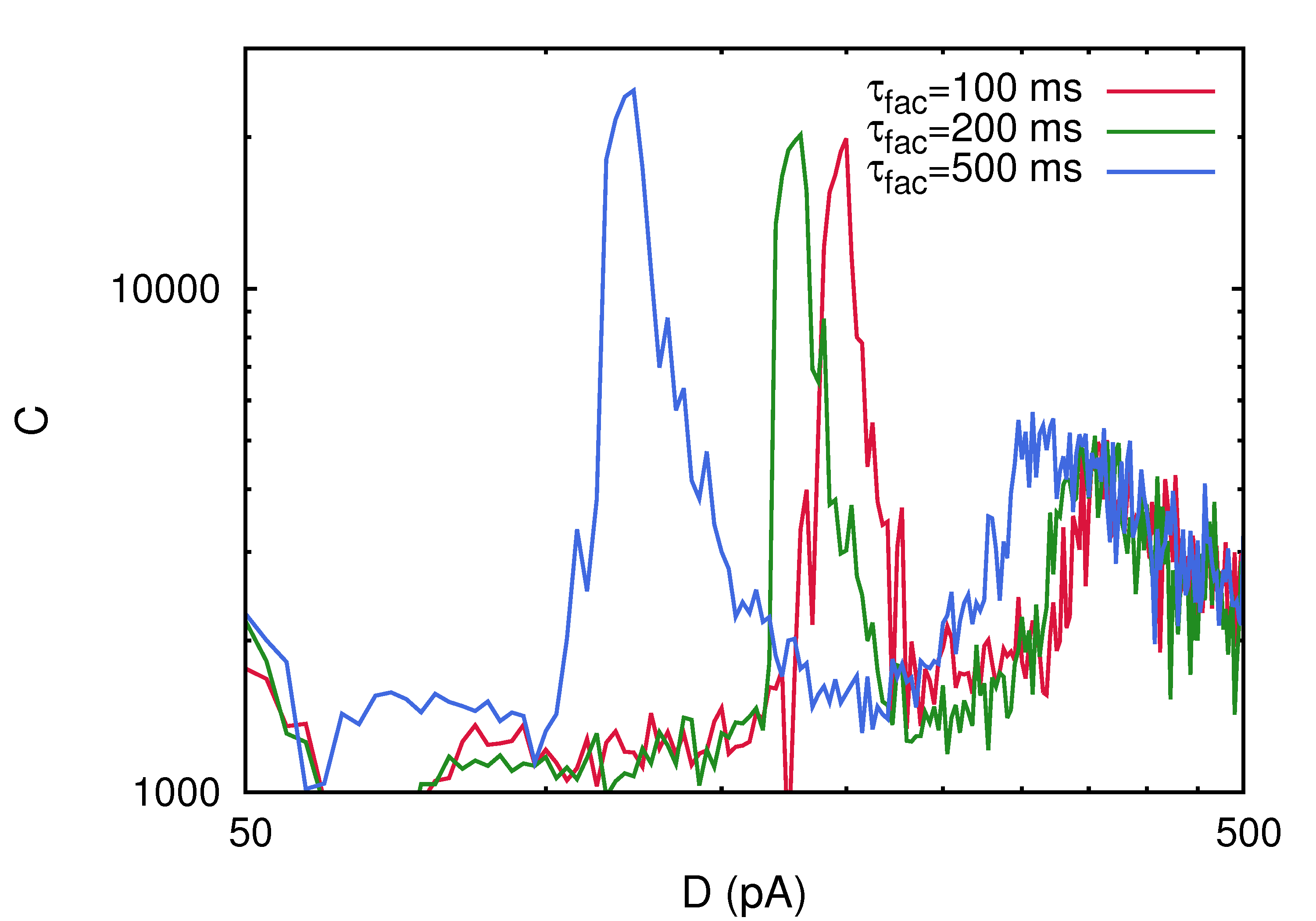} 

{\Large Figure 5} 
\end{center}

\newpage

\begin{center}
\includegraphics{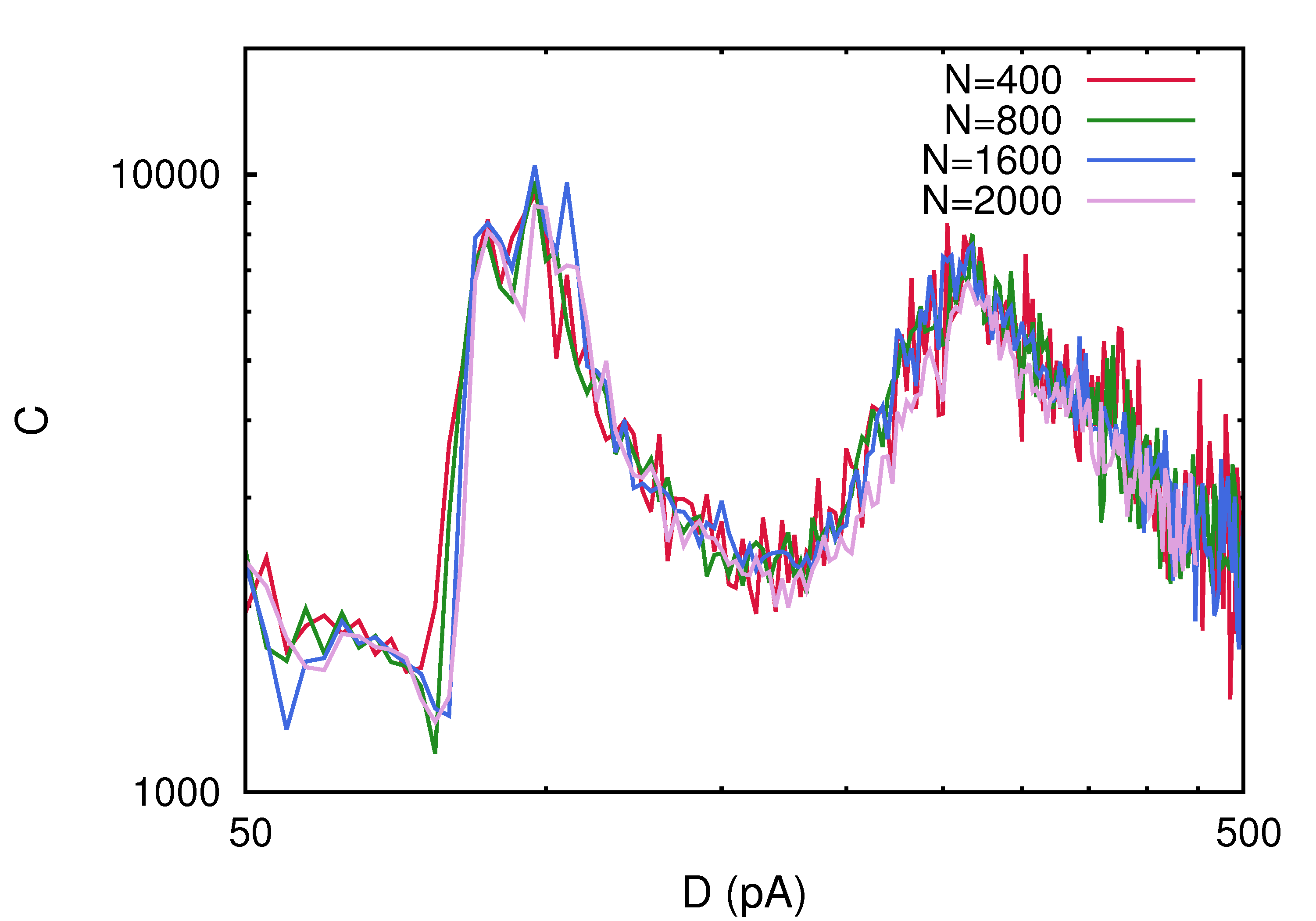} 

{\Large Figure 6} 

\end{center}

\newpage

\begin{center}
\includegraphics{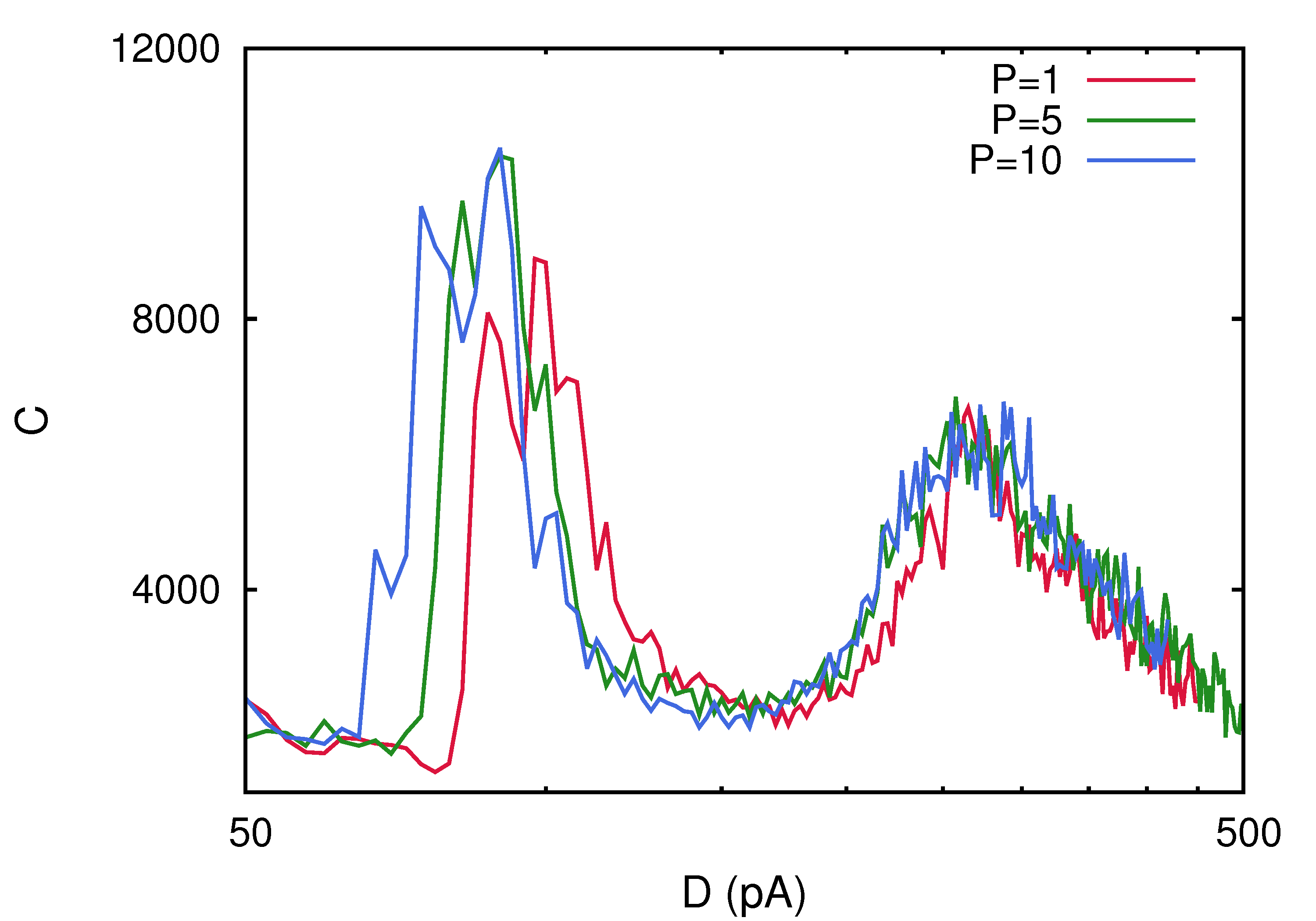} 

{\Large Figure 7}

\end{center}

\newpage

\begin{center}
\includegraphics[scale=0.5]{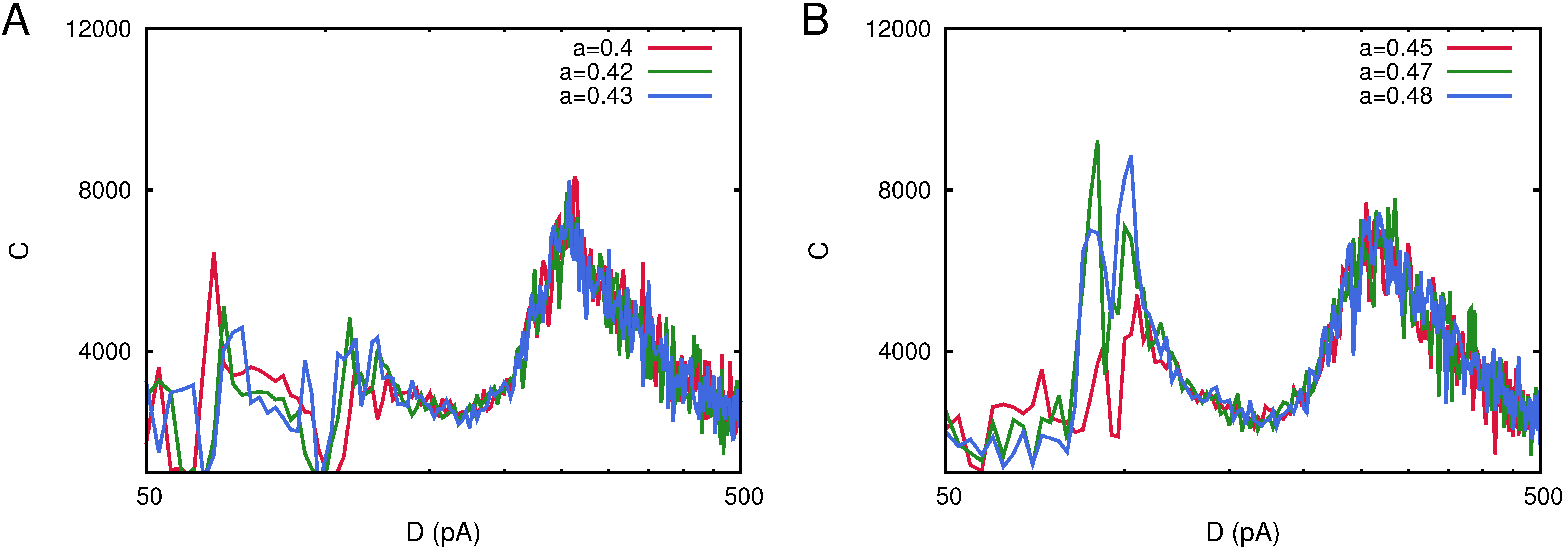} 

{\Large Figure 8} 

\end{center}

\newpage

\begin{center}
\includegraphics[scale=0.5]{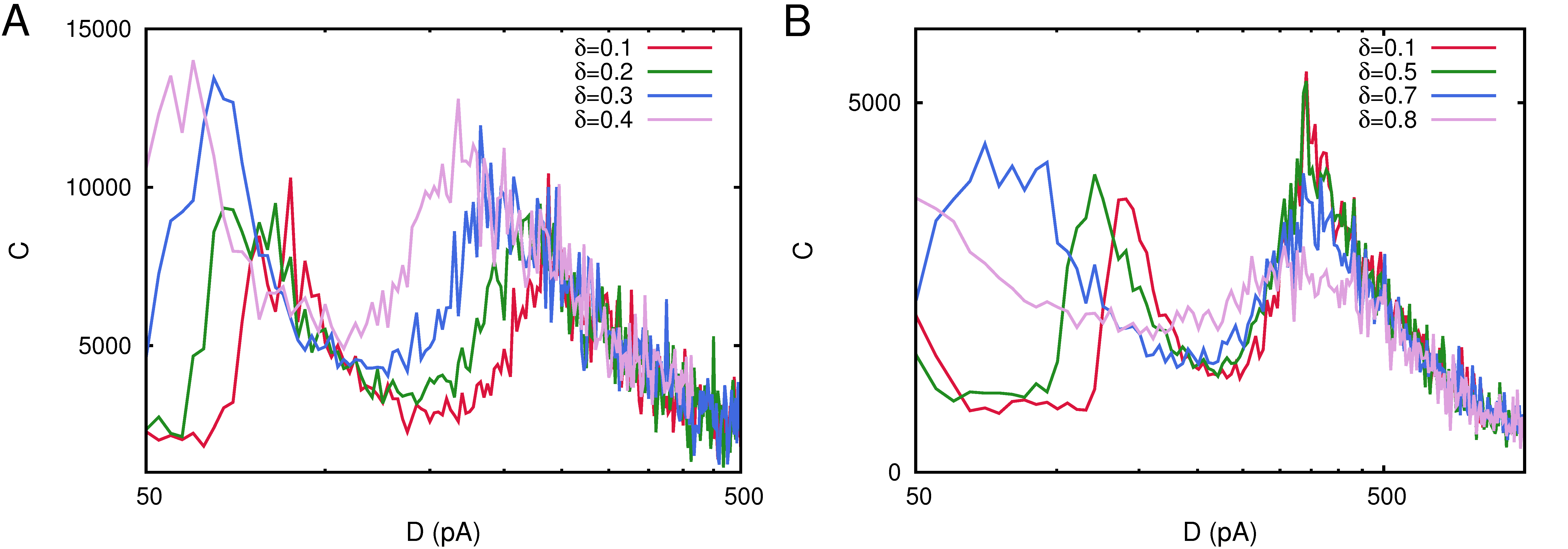}

{\Large Figure 9}

\end{center}

\newpage

\begin{center}

\includegraphics[scale=0.5]{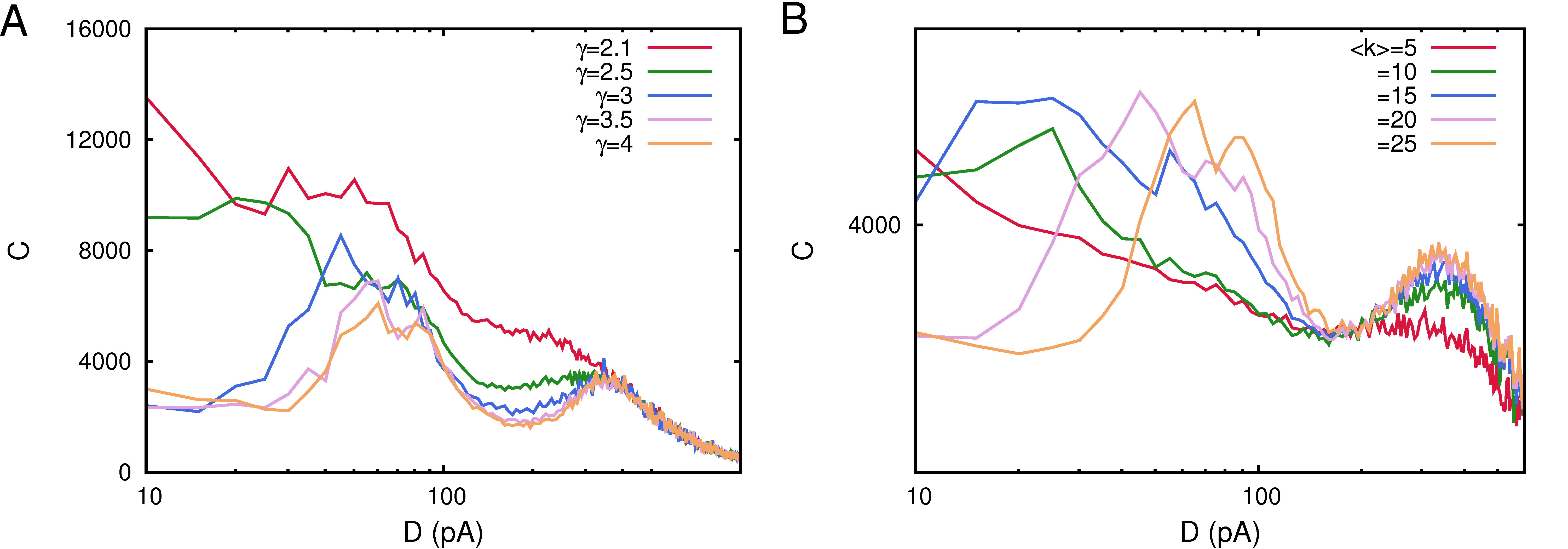} 

{\Large Figure 10} 
\end{center}

\newpage

\begin{center}
\includegraphics[scale=0.7]{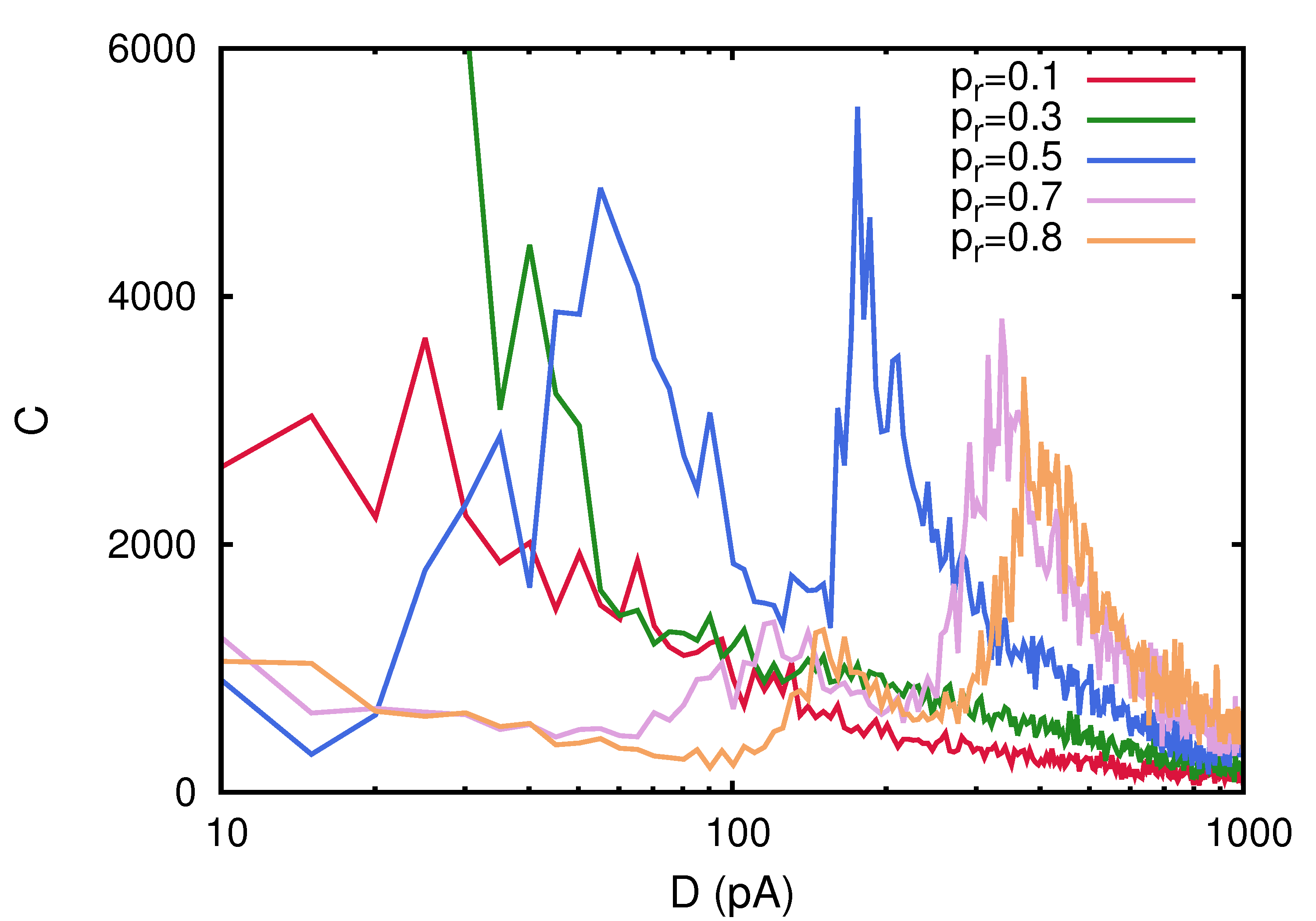} 

{\Large Figure 11} 
\end{center}

\newpage

\begin{center}
\includegraphics[scale=0.7]{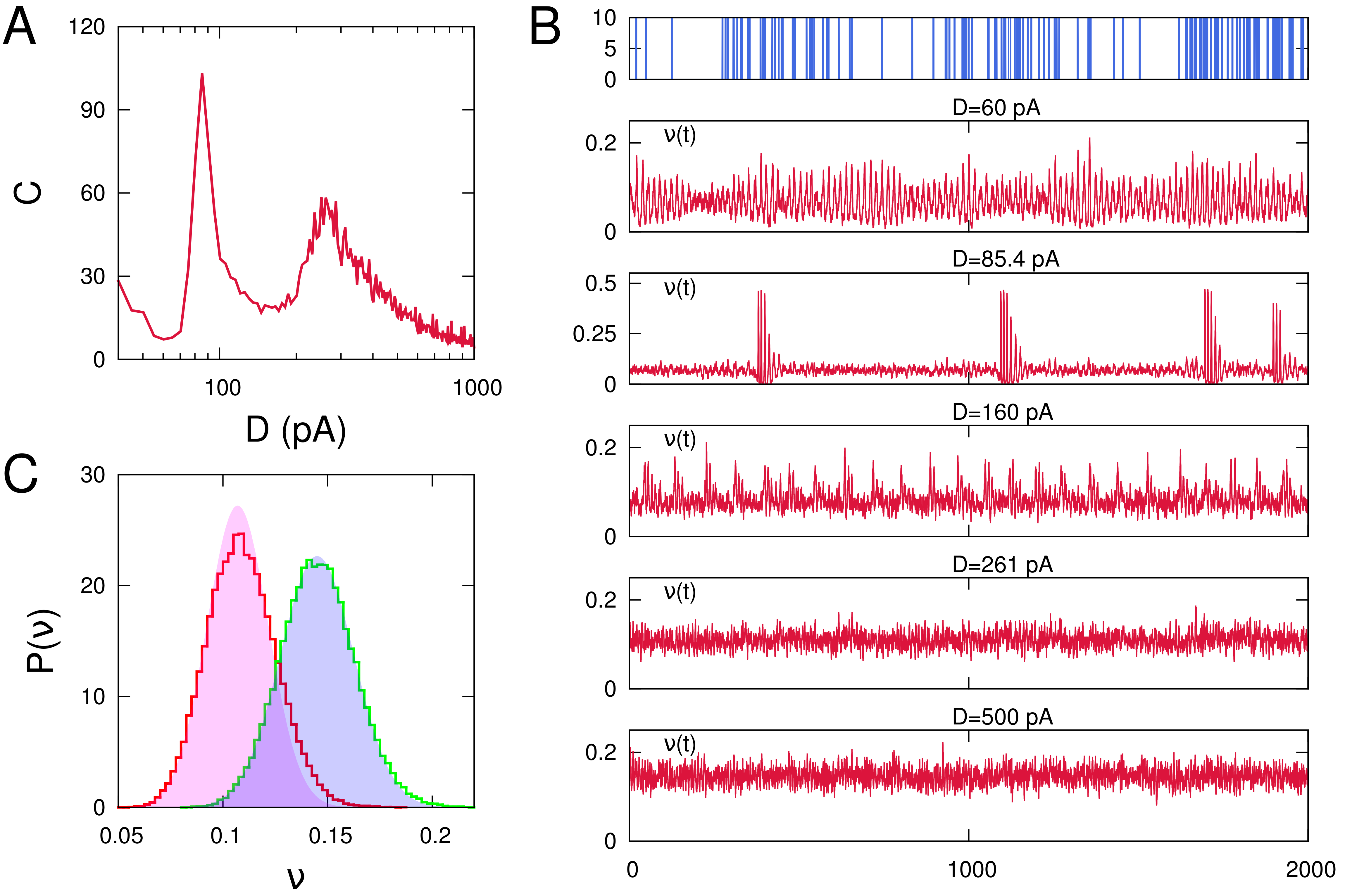} 

{\Large Figure 12} 
\end{center}

\newpage

\begin{center}
\includegraphics{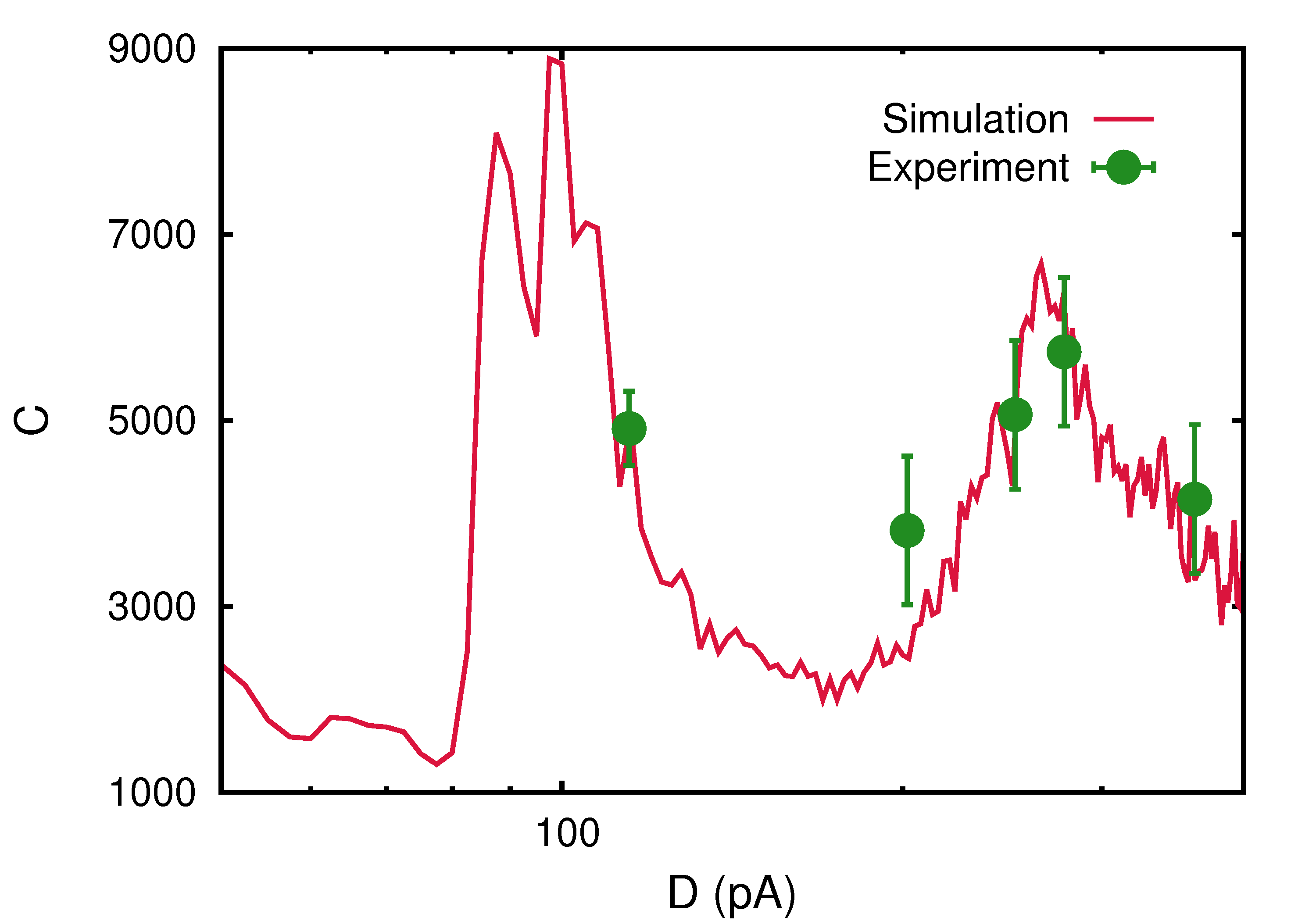} 

{\Large Figure 13} 

\end{center}

\newpage

\end{document}